\documentclass[journal,twocolumn]{IEEEtran}
\usepackage{cite}
\usepackage{amsmath,amssymb,amsfonts}
\usepackage{tcolorbox}
\usepackage{algorithmic}
\usepackage{graphicx}
\usepackage{textcomp}
\usepackage{xcolor}
\usepackage{array}
\usepackage{float}
\usepackage{url}
\usepackage{fancyvrb}
\usepackage{booktabs}
\usepackage{mathtools}

\usepackage{hyphenat}
\usepackage{newfloat}
\DeclareFloatingEnvironment[fileext=frm,listname={List of models},placement={!ht},name=Model]{modelenv}

\newtcolorbox{modelbox}{colback=white, 
 colframe=black,
 width=\textwidth,
 top=-1mm, 
 bottom=-0mm
}

\newcommand{\model}[3]{
\begin{modelenv}
\begin{center}
\begin{minipage}[H]{0.45\textwidth}
	\begin{small}
		\caption{#1}
		\label{model:#2}
	\end{small}
	\begin{modelbox}
	\begin{small}
		\smallskip\noindent
		\begin{flalign*}
			 #3
		\end{flalign*}
	\end{small}
	\end{modelbox}
\end{minipage}
\end{center}
\end{modelenv}
}

\graphicspath{ {./figures/} }

\usepackage{hyperref}
\hypersetup{
    colorlinks=false,
    pdftitle={Statistical Models for the Analysis of Optimization Algorithms with Benchmark Functions},
    pdfauthor={D.I. Mattos, J. Bosch, H.H. Olsson}
}

\begin{document}

\title{Statistical Models for the Analysis of Optimization Algorithms with Benchmark Functions}

\author{David Issa Mattos,
        Jan Bosch,
        Helena Holmström Olsson
\thanks{D. I. Mattos and J. Bosch are with the Department of Computer Science and Engineering at Chalmers University of Technology, Gothenburg, Sweden}
\thanks{H. H. Olsson is with the Department of Computer Science and Media Technology at Malmö University, Malmö, Sweden}
}


\maketitle

\begin{abstract}
Frequentist statistical methods, such as hypothesis testing, are standard practices in studies that provide benchmark comparisons.  Unfortunately, these methods have often been misused, e.g., without testing for their statistical test assumptions or without controlling for family-wise errors in multiple group comparisons, among several other problems. Bayesian Data Analysis (BDA) addresses many of the previously mentioned shortcomings but its use is not widely spread in the analysis of empirical data in the evolutionary computing community. This paper provides three main contributions. First, we motivate the need for utilizing Bayesian data analysis and provide an overview of this topic. Second, we discuss the practical aspects of BDA to ensure that our models are valid and the results are transparent. Finally, we provide five statistical models that can be used to answer multiple research questions. The online appendix provides a step-by-step guide on how to perform the analysis of the models discussed in this paper, including the code for the statistical models, the data transformations, and the discussed tables and figures. 
\end{abstract}


\section{Introduction}
\label{sec:introduction}

With the increasing number of optimization algorithms being developed, benchmarks are used not only to show that the new algorithms work but also to compare algorithms against each other \cite{hansen2020coco, nevergrad, li2013benchmark, Jamil2013}.

Over the years, algorithms for black-box optimization have been demonstrated to work and have been compared with each other using simple descriptive statistics (such as mean, standard deviation, and median), boxplots and performance profiles \cite{dolan2002benchmarking}. Recently, frequentist statistical methods such as null hypothesis testing have become standard practice in papers that provide benchmark comparisons \cite{barr1995designing, garcia2009study, lacoste2012bayesian}. 

Unfortunately, frequentist methods for null hypothesis testing have often been misused by scientists and practitioners looking for a dichotomy tool to assess a particular problem, without an evaluation of the size of the observed effect or a discussion with complementary analysis\cite{wasserstein2019moving}. By utilizing statistical tests as black-box tools, many pitfalls and misuses have been observed in different fields of science. We list some of the observed pitfalls.
(1) Lack of separation between the effect size and sample size in the p-value \cite{benavoli2017time};  
(2) Lack of information regarding the null hypothesis \cite{benavoli2017time, kruschke2018bayesian, gill1999insignificance}; 
(3) Misinterpretation of the actual meaning of the p-value (including by instructors in statistics) \cite{haller2002misinterpretations, wasserstein2016asa, gill1999insignificance }; 
(4) Misinterpretation of the meaning of confidence intervals \cite{cumming2013understanding, furia2019bayesian}; 
(5) Lack of transparency in the reporting of the statistical procedures (such as providing the value of test statistics, the actual value of the p-value, confidence intervals) \cite{cumming2013understanding}; 
(5) Common problems related to the misuse of the statistical tests such as not verifying the statistical test assumptions, not controlling for correlated samples or not controlling for family-wise errors in multiple group comparisons \cite{wilcox2011modern}.

Bayesian Data Analysis (BDA) \cite{gelman2013bayesian} has gained attention as a potential replacement for frequentist statistics by providing an easier to interpret data analysis framework and by addressing many of the previously mentioned shortcomings.  Despite the popularity of Bayesian-based optimization algorithms, BDA has not been widely used for the analysis of benchmark data in Evolutionary Computing. With this paper, we argue for the adoption of BDA in evolutionary computing, in particular in the analysis of benchmark data. Specifically, we provide an overview of this topic discussing practical aspects of BDA to ensure that our models are valid and the results are transparent to avoid misuse and pitfalls. Focusing on the interpretation and answering specific research questions, we provide five statistical models together with a reproducible appendix that contains a step-by-step analysis, including the code for the statistical models, the data transformations, and the discussed tables and figures. We reinforce that BDA is not a black-box framework for decisions. BDA requires convergence checking, validation of the statistical models, reporting transparency and it is complementary with other types of analysis such as plots and tables. A correct interpretation of the results still requires domain specific knowledge and understanding of the statistical models and their limitations.

The remainder of this paper is organized as follows. Section \ref{sec:relatedwork} discusses related work. Section \ref{sec:bda} provides an overview of Bayesian data analysis, including a discussion of practical aspects of Bayesian data analysis to ensure that our models are valid and the results transparent. Section \ref{sec:empiricaldata} provides a description of the empirical data used to exemplify all the statistical models. Section \ref{sec:statsmodels} provides a discussion of each statistical model, interpretation and presentation of the results. Section \ref{sec:conclusion} concludes this paper.

The online appendix can be found at: \url{https://doi.org/10.5281/zenodo.4067712}
\section{Related work}
\label{sec:relatedwork}

In this section, we provide an overview of related works in statistical analysis for the development and comparison of algorithms.  

In the frequentist setting, the work by  Eftimov et al. \cite{eftimov2018impact,eftimov2017novel} provides an analysis of different ranking schemes for analysis of benchmarking in evolutionary computation, concluding that different statistical tests can lead to different ranking schemes. To overcome this, they propose the use of a different scheme that uses the whole distribution instead of only the average or the median and a new analysis method. The proposed approach is shown to be more robust to outliers and the ranking scheme. However, the proposed approach is used only in ranking situations and it does not take into account additional covariates and the internal correlation between the data given by the benchmark functions. Despite the focus on the practical significance, it is still subjected to the other problems of frequentist analysis \cite{furia2019bayesian}. 

Gagliolo and Legrand \cite{gagliolo2010algorithm} provide an overview of the survival analysis to runtime distributions, discussing its usage in the algorithm selection. The paper discusses the application and interpretation of the frequentist variation of survival models such as the parametric Cox Proportional Hazard model and the non-parametric Kaplan-Meyer estimator. In Section \ref{sec:timetoconverge}. we discuss survival analysis and its application in the context of Bayesian Data analysis. Chiarandini and Goegebeur \cite{chiarandini2010mixed}  discuss the use of frequentist linear and multilevel and models for different experimental designs. They reinforce the need to separate the effects of the algorithms from the problem instances by modeling the instances with random factors. All the statistical models in Section \ref{sec:statsmodels} utilize random factors to separate the effects of the benchmarks from the algorithms.

Bartz-Beielstein et al. \cite{bartz2020benchmarking} provide an extensive survey that discusses different topics for promoting good benchmark practices, from objective statement and selection of problems to the analysis and presentation of results. However, from the analysis perspective, the paper focuses solely on the use of frequentist statistics, while the known limitations and pitfalls of frequentist statistics are not considered and BDA is not mentioned.

In the Bayesian setting, the work by Calvo et al. \cite{calvo2019bayesian} provides the first paper for Bayesian estimation in evolutionary computing. In the paper, they provide a practical application of Bayesian data analysis on the comparison of eleven algorithms on 23 optimization problems. The authors discuss the Plackett-Luce model for ranking algorithms and compare it with the frequentist approach. The algorithms are analyzed in stratified benchmark functions (e.g. easy, medium, hard). In our work, we discuss an extension of the Bradley-Terry model that, under some conditions, is equivalent to the Plackett-Luce model for complete ranks but we also consider the random effects of the benchmarks.  Furia et al. \cite{furia2019bayesian}  provide a discussion of frequentist and Bayesian data analysis in empirical software engineering (including a re-analysis of early research). They discuss many of the shortcomings of frequentist statistics and provides an introduction to Bayesian data analysis from the Software Engineering perspective.

Carrasco et al. \cite{carrasco2017rnpbst} discuss the application of both frequentist and Bayesian non-parametric tests in the comparison of machine learning algorithms. They discuss three Bayesian tests, a variation of the t-test that takes into account correlation in the results, and two nonparametric tests. The discussed Bayesian tests are based on the work of Benavoli et al. \cite{corani2015bayesian, benavoli2014bayesian} that provide closed formulas for the nonparametric tests under specific prior conditions. The work by Lacoste et al. \cite{lacoste2012bayesian} discusses a Poisson binomial test to compare algorithms and demonstrates that the approach is more reliable than the sign test and the Wilcoxon signed-rank test. 

In our work, we base our analysis on parametric models that are of interest for ``scientific experimental analysis, where the interest is in explaining the causes of the success of a certain optimization approach rather than in mere comparative studies'' \cite{chiarandini2010mixed}.  At the expense of computational resources, our models can utilize flexible priors since the posterior is computed numerically (instead of analytically) by a Markov Chain Monte Carlo (MCMC) sampler. Additionally, our models  take into account the correlation by the benchmarks and can be easily extended to additional correlations either nested or on the same level.

\section{Bayesian Data Analysis}
\label{sec:bda}

In this section, we provide a short overview of the Bayesian data analysis process and some practical aspects to ensure that our models are valid and our results transparent. A full comparison between the Bayesian and the frequentist framework is beyond the scope of this paper and we refer to other sources \cite{furia2019bayesian, mcelreath2020statistical, kruschke2013bayesian,kruschke2018bayesian}.

The main idea behind Bayesian data analysis is the reallocation of credibility across possibilities \cite{kruschke2018bayesian}. In practical terms, this means that we start with a prior explanation of the results before seeing any data and a model on how the data is generated. As we collect new data, our beliefs about the system are reallocated. The probability of candidate explanations that do not fit well the data is therefore reduced. In this updating process, we get a probability distribution of each possible explanation of the data. This allows us to obtain the credible (or uncertain) intervals \cite{gelman2013bayesian, furia2019bayesian}.

The process of allocating explanations into probability distributions happens through the principles of conditional probabilities and the Bayes theorem \cite{furia2019bayesian}:

\begin{equation}
\mathcal{P}(h|d) = \dfrac{\mathcal{P}(d|h)\cdot \mathcal{P}(h)}{\mathcal{P}(d)},
\end{equation}

\noindent
where $d$ represents the data, $h$ the explanation (or hypothesis), $\mathcal{P}(h|d)$ is the conditional probability of the hypothesis given the observed data. 
Below are common names for the factors in the Bayes theorem:
\begin{itemize}
    \item $\mathcal{P}(d|h)$ is called the likelihood of the data $d$ under the hypothesis $h$.
    \item $\mathcal{P}(h)$ is called the prior.
    \item $\mathcal{P}(h|d)$ is called the posterior. The posterior represents the probability distribution of each parameter estimate (the hypothesis $h$) given the observed data
    \item  $\mathcal{P}(d)$ is called the marginal likelihood and it is a constant, that is often impossible to compute analytically.
\end{itemize}

\subsection{Bayesian tools}
Several tools are capable of performing Bayesian data analysis, such as IBM SPSS, SAS, Stata, JASP, R, TensorFlow, Stan among others. Although the discussed models can be used in most statistical software, we utilize the Stan software and its modeling language \cite{carpenter2017stan}, while we utilize the R language for data transformation and plotting. Stan can be easily integrated with R, with the \texttt{rstan} package\footnote{\url{https://cloud.r-project.org/package=rstan}}.  Many R packages make modeling in Stan easier (e.g. \texttt{rstanarm}\footnote{\url{https://cloud.r-project.org/package=rstanarm}} and \texttt{brms} \footnote{\url{https://cloud.r-project.org/package=brms}}). However, in the online appendix, we decided to provide the raw Stan model, since it can be used together with many different programming languages such as  Python, Stata, Julia,  Matlab, where readers can perform the data transformation and plotting in their preferred language while utilizing the same statistical models that we provide.

\subsection{Bayesian inference and MCMC}
While the Bayes theorem provides how our initial beliefs are going to be updated based on the observed data to generate the posterior, the actual computational process is more complicated due to the marginal likelihood. The posterior can be approximated without explicitly computing  $\mathcal{P}(d)$  by using a class of sampling algorithms called Markov Chain Monte Carlo (MCMC). The goal of an MCMC is to generate an accurate representation of the posterior of the model parameters \cite{kruschke2013bayesian} by generating a large representative sample of credible intervals that represents the posterior distribution \cite{kruschke2018bayesian}. Although there are several MCMC algorithms, we utilize in this work the Hamiltonian MCMC No U-Turn algorithm (NUTS) \cite{hoffman2014no} available in the Stan program \cite{carpenter2017stan}, as it provides faster convergence, handles correlated parameters in the posterior better than others, and provides good diagnostic tools when the chain diverges.

One of the main disadvantages of Bayesian data analysis is the time necessary to compute the posteriors of an MCMC process compared to maximum likelihood estimates from the frequentist approach. 

\subsection{Posterior and intervals}
One of the benefits of the Bayesian approach is that we get a posterior distribution of the estimated parameters from a model. With the posterior, we can get not only point estimates (such as the mean or median of a parameter) but also credible intervals (also called uncertain intervals) of these parameters. These intervals are useful to estimate the uncertainty of the parameters without making assumptions of repeated sampling, such as the confidence intervals given by the frequentist approach (that assumes that a fixed point estimated will lie within an interval if the sampling process is repeated many times and assuming that the null hypothesis is true). The credible interval is a probabilistic statement about the real value of a parameter while the confidence interval reveals only the uncertainty about the interval (if it contains the value or not). Confidence intervals, due to their assumptions, cannot be used to understand the probability of a point estimate parameter.

In Bayesian data analysis, we make use of the full posterior of the parameter to make interval inferences instead of single-point estimates. To help our analysis, three common intervals are used:

\subsubsection{Equal tail interval} this is a credible interval that divides the posterior lower and higher tails equally, based on quantiles. For example, the 95\% interval will exclude 2.5\% of the data in each tail. This is the default interval that Stan provides after sampling in R.

\subsubsection{Highest Posterior Density  (HPD) Interval} this is the narrowest interval in a unimodal distribution that will contain the specified probability mass, the area under a density distribution.  This is the interval that best represents the parameter values consistent with the data \cite{mcelreath2020statistical}. Throughout this paper, we will present the HPD intervals for the parameters we estimate.

\subsubsection{Region of Practical Equivalence (ROPE)} is a practical interval that encloses the values that are considered negligible from a practical perspective \cite{kruschke2013bayesian}. This interval combined with a credible interval like the HPD interval can be used for decisions. For example, if from a practical perspective an improvement or degradation of an algorithm of $x_0 \pm 10\%$ around a baseline $x_0$ is considered irrelevant, this is the ROPE interval. If all or almost all of the HPD interval (given a threshold such as 95\% of the interval) falls in the ROPE interval we can say that the algorithm doesn't provide practical improvement. If the HPD interval falls above or below the ROPE interval we can say that there is a real improvement or degradation respectively. If there is a large overlap, we cannot make an accept or reject statement like that. Note that different to the frequentist approach, with such tools you can accept the null hypothesis or reject it and do so without being concerned with one or two-tail hypothesis and the family-wise error. It is worth noting that the ROPE interval is highly dependent on the practical values that determine it and the scale of the parameters (e.g. they are at different scales in a binomial and a normal regression).

To avoid misuse of this interval and make mistakes similar to the ones often seen in the null hypothesis significant testing, the use and reporting of the ROPE interval should be explicitly specified and justified. Otherwise, it is preferred that it remains unspecified to allow readers to use and assess the results with their own ROPE intervals \cite{kruschke2013bayesian}. In Section \ref{sec:statsmodels}, we opted to omit the ROPE intervals.

\subsection{Model checking}
Since we can get inference from prior-to-posterior on any reasonable model we should perform additional model checking procedures to ensure the validity of the models. Therefore, in a good Bayesian data analysis we should check for proper convergence of the MCMC, for the adequacy of the model fit with the data, and check for the model robustness against different modeling choices \cite{gelman2013bayesian}.

\subsubsection{Sampling convergence}
After specifying the model, we need to specify the sampling parameters and assess the convergence of chains for valid inference of the posterior. To allow diagnostics of the sampling process, we follow the recommendations of the Stan software\footnote{\url{https://mc-stan.org/users/documentation/}} and initialize the sampling with four chains, random initial values and target Metropolis acceptance rate equals to 0.8. For the number of iterations and warmup, we adjust accordingly to the number of effective samples of the posterior and whether there are divergent iterations.

\textbf{Trace plots of the chains: } These are diagnostic plots to look at the sampling of each chain. All chains should be well-mixed without any pattern or trend \cite{mcelreath2020statistical}.

\textbf{Number of effective samples of the posterior $n_{\text{eff}}$: } Markov Chains are typically autocorrelated, which will result in the autocorrelation of the samples. The number of effective samples of the posterior indicates the number of independent samples. Stan provides warning messages if there is a low number of effective samples in the posterior. As a rule of thumb \cite{mcelreath2020statistical}, 200 effective samples of the posterior are enough to estimate the mean of a parameter but we might require more if estimating quantiles or highly skewed posteriors.

\textbf{Gelman-Rubin potential scale reduction  ($\hat{R}$): } this statistic measures the ratio of the average variances of samples within each chain to the variance across chains \cite{gelman1992inference}. This is a parameter that indicates the convergence of the chains. If the chains have not converged, the $\hat{R}$ will be greater than one. In practical terms, we require values of $\hat{R}<1.05$ and preferably $\hat{R}<1.01$ \cite{mcelreath2020statistical}

\textbf{Number of divergent iterations: } During the sampling procedure we specify a warmup period in which the sampler learns which parameters to use. During this period, we can have divergent iterations. However, after warmup, there should be no divergent iterations. If there are, the posterior estimates cannot be considered valid.

\subsubsection{Choice of priors}
One aspect commonly discussed and criticized in BDA is the subjectiveness of the priors. Priors are part of the modeling flexibility that BDA provides to researchers. It adds the possibility of incorporating prior knowledge of previous research on the model to create better and more robust models. For example, a prior indicating that a parameter should be within a range of -10 to +10 might be added o constraint the parameter value. This is often a more reasonable approach for the vast majority of cases than allowing a parameter to have a range between $-\infty$ to $+\infty$ (as the frequentist data analysis does). 

This flexibility also allows researchers to choose between \textit{non-informative}, \textit{weakly informative}, and \textit{informative} priors for their models. A non-informative prior is based on bounded or unbounded uniform distribution and does not aggregate any information to the posterior. 

Weakly-informative priors are those that do not impact or aggregate much information in the posterior parameters but it is not as vague as the non-informative prior. They can act as regularizing priors and improve the inference and convergence of the MCMC \cite{mcelreath2020statistical}. An example of such a prior would be a normal distribution with a large variance compared to the expected parameter value. 

Informative priors are those that incorporate previous knowledge on the subject to improve the model. These priors impact the posterior parameters. As a rule of thumb\footnote{\url{https://github.com/stan-dev/stan/wiki/Prior-Choice-Recommendations\#how-informative-is-the-prior}}, if the posterior standard deviation of a parameter is more than 0.1 times the prior standard deviation, the prior is considered informative. A classification of the priors in informative and weakly informative is not only a matter of the prior distribution (and its parameters) but the joint effect of the prior, the number of parameters in the model, and the amount of collected data. For small datasets, the prior will have a larger influence in the parameter estimate compared to larger datasets. Therefore adjusting the prior distribution parameters to be weakly-informative should be done together with the actual data and model in question.

For all the models in Section \ref{sec:statsmodels}, we adjusted the priors and hyperpriors to be weakly-informative priors based on the presented rule of thumb.

\subsubsection{Model comparison}
\label{sec:modelcomparison}
For the same data, we might have different valid model candidates (including different priors and likelihoods), and we should compare these models and verify their performance. A recommended approach is to start with simple models and start building more complex models. If the complexity does not increase the predictive accuracy and it is not justifiable theoretically, simpler models are preferred. Comparing the predictive accuracy can be done by analyzing the model entropy information with the Watanabe-Akaike Information Criteria WAIC \cite{gelman2014understanding} or the Leave-One-Out Cross Validation method (LOO-CV)\cite{gelman2013bayesian}. The calculation of the WAIC and the LOO-CV requires the log-likelihood, which is not calculated automatically in Stan. However, the models we use, and available in the repository, calculate this value.

Note that, one should not use the AIC or BIC criteria in the Bayesian context since both methods assume that the model utilizes flat priors and the maximum a posteriori estimate \cite{mcelreath2020statistical}. These assumptions are often not true since flat priors are discouraged and the estimation method is the MCMC. Information criteria such as the WAIC provide results equivalent to the AIC (when assumptions are met) and can also be used under different priors and estimation methods \cite{mcelreath2020statistical}.

\subsubsection{Sensitivity analysis}
\label{sec:sensitivity}
During the development of a model, many alternative models might be considered, including different choices of likelihood, prior, predictors, etc. Sensitivity analysis is a process to evaluate how much the posterior inferences change when we change different aspects of the model. For example, we might have different choices of priors. With sensitivity analysis, we can evaluate the impact of these priors in the inferences we make. If after the modifications, the inference results remain unchanged, we can say that the posterior inferences are robust. We can perform a sensitivity analysis directly on the posterior parameters, verifying if they still have similar magnitudes and directions, but also in terms of their posterior predictive checks. Sensitivity analysis overcomes the critique of the subjectiveness of BDA with transparency \cite{mcelreath2020statistical}. We provide an example of sensitivity analysis in the online appendix.

\subsubsection{Posterior predictive checking}
Finally, the last step to analyze the validity of a model is through a posterior predictive check \cite{gelman2013bayesian}. Posterior predictive checking is a way to assess how large are the residuals of the model, i.e. the difference between the predictive values of the model compared to the observed values. If the model does not predict or can't explain the data well, it might not be a good or even valid model.

\subsubsection{Sample size and power analysis}
In BDA analysis, it is also possible to perform both prospective and retrospective power analysis without some of the criticisms of frequentist power analysis \cite{miller2009probability, kruschke2013bayesian}. Prospective power analysis (such as determine an ideal sample size) consists of generating point estimate hypothesized parameter values for the model and using this model to generate data with different sample sizes \cite{kruschke2013bayesian}. The generated data is used to estimate the posterior parameters of the model. If the HPD interval is contained inside a ROPE interval (that specifies the desired size of the uncertainty interval), for each parameter estimate,  we consider that we have satisfied the uncertainty interval restriction with an appropriated sample size. Since this study does not place a restriction on the size of the uncertainty intervals for each model, we do not perform a sample size analysis to collect the data. Therefore, our estimates and uncertainties are in accordance with the data collected as described in Section \ref{sec:empiricaldata}.

\section{The empirical data}
\label{sec:empiricaldata}
In this section, we first present an overview of the algorithms and the benchmark functions that are used to collect the empirical data. Then we present the research questions that guided the development of the models in Section \ref{sec:statsmodels}.

Since the goal of this work is to illustrate the statistical models and not to provide an in-depth comparison of the state-of-the-art algorithms, we created a simplified experimental simulation scenario with enough complexity to fully illustrate the statistical methods.  We performed an empirical evaluation of eight well-known algorithms for black-box optimization against 30 benchmark functions under different noise and budget conditions.  In this experimental simulation, we focus only on continuous benchmark functions, although the discussed models can be applied and extended to other problems.

\subsection{The algorithms}
The choice of the first six algorithms is based on their widespread use, the computational speed (CPU time) for each function evaluation, and the easy availability of a Python 3 implementation, on which the simulation framework is based. Also, we selected a random search algorithm to be used as a baseline in some of the comparisons, and a variation of the random search (Random Search x2).  The Random Search x2 can be used as a baseline for the comparison in the cases in which there is noise in the output value of the benchmark function.

We utilize the following algorithms with their respective default parameters from the software package:
\begin{itemize}
     \item \textbf{Particle Swarm Optimization} \cite{kennedy1995particle}. We utilize the implementation from the \texttt{NiaPy} package \cite{NiaPyJOSS2018} with the following parameters: $C_1=2$ (cognitive component) , $C_2=2$ (social component), $w=0.7$ (inertial weight), $v_{min}=-1.5$ (minimal velocity), $v_{max}=1.5$ (maximal velocity) and population of 30.
    \item \textbf{Cuckoo Search} \cite{yang2009cuckoo}. We utilize the implementation from the \texttt{NiaPy} package \cite{NiaPyJOSS2018} with the following parameters: $p_a=0.2$ (proportion of worst nests) and $\alpha=0.5$ (scale factor for the Levy flight) and population size of 30.
    \item \textbf{Simulated Annealing} \cite{kirkpatrick1983optimization}. We utilize the implementation from the \texttt{NiaPy} package \cite{NiaPyJOSS2018} with the following parameters: $\delta =0.5$ (movement of neighbor search, $T=2000$ (starting temperature), $\Delta _T = 0.8$ (change in temperature), $\epsilon =1e-23$ (error value) and a linear cooling method.
    \item \textbf{Differential evolution} \cite{storn1997differential}. We utilize the implementation from the \texttt{NiaPy} package \cite{NiaPyJOSS2018} with the following parameters: $F=1$ (scale factor), CR$=0.8$ (crossover probability), random cross mutation  and population size of 30.
    \item \textbf{Nelder-Mead} \cite{lagarias1998convergence}. We utilize the implementation from the \texttt{NiaPy} package \cite{NiaPyJOSS2018} with the following parameters: $\alpha=0.1$ (reflection coefficient), $\gamma=0.3$ (expansion coefficient), $\rho=-0.2$ (contraction coefficient) and $\sigma=-0.2$ (shrink coefficient).
    \item \textbf{Covariance Matrix Adaptation Evolution Strategy} (CMA-ES) \cite{hansen2006cma}. We utilize the implementation from the package \texttt{pycma} \cite{hansen2019pycma}. We utilize the following parameters $\sigma_0=0.5$ (initial standard deviation).
    \item \textbf{RandomSearch1} (Random Search x1) Random samples are selected to search the space. Each sample is evaluated only once. We utilized our own implementation.
    \item \textbf{RandomSearch2} (Random Search x2) is a variation of Random Search in which the same point in the search space is evaluated twice before getting a new sample. We utilized our own implementation.
\end{itemize}

\subsection{The benchmark functions}
For the benchmark functions, we randomly selected 30 benchmark functions from a pool of 220 benchmark functions from both the BBOB-2009 \cite{Finck2010} function definitions as well as from a literature survey \cite{Jamil2013}. All the benchmark functions are used for the optimization of continuous parameters. To ensure that they were correctly specified, they were tested nightly, for over a month, with a random search algorithm to verify whether all the global minima were identified and if these were indeed the global minima. Additionally, due to the computational time required, we restricted the benchmark functions to a limit of 6 dimensions, which is sufficient to illustrate the generality of the statistical models.

For a mathematical definition of these functions, we refer to the survey \cite{Jamil2013} and  the BBOB-2009 function definitions \cite{Finck2010}. The used benchmark functions are: Sphere 6-D, Tripod, ChungReynolds 2-D, Pinter 6D, StrechedVSineWave 2-D, Trigonometric-1 6-D, BentCigar 6-D,  ChenV, Discus 2-D, Schwefel2d20 2-D, ChenBird, Schwefel2d21 6-D, Zakharov 2-D, Damavandi, Schwefel2d4 6-D, Whitley 6-D, Shubert, XinSheYang2 2-D, Mishra7 6-D, Schwefel2d23 6-D, Exponential 2-D, Salomon 2-D, RosenbrockRotated 6-D, Qing 2-D, LunacekBiRastrigin 6-D, ThreeHumpCamelBack, Schwefel2d26 6-D, Trefethen, Price1 and Giunta.

\subsection{The experimental conditions}
All algorithms were run for all benchmark function ten times in each combination of an experimental condition:
\begin{itemize}
    \item \textbf{Noise}: 0 or 3.0. When noise is present, we added a Gaussian random variable on the output of the benchmark function with a mean in the benchmark value and a standard deviation of 3.0. Although this might impact differently each benchmark function result, it represents a constant measurement error in real-world conditions.
    \item \textbf{Budget}: $20$, $10^2$, $10^3$, $10^4$ and $10^5$  function evaluations per number of dimensions of the benchmark function.
\end{itemize}

These conditions resulted in a total of $24,000$ data points in our data set, in which each point corresponds to one algorithm run.

\subsection{The logged metrics}

Apart from the benchmark functions metrics (e.g. number of dimensions) and the experimental conditions (noise level, budget), we logged additional metrics for each algorithm. For these metrics we use the following notation:

$\mathbf{X} \in \chi$  from a compact subset space $\chi \subset \mathbb{R} ^d$ is a vector of the input for the benchmark function and has dimension $d$ .

$\mathbf{X^*} $ is the global minimum of the benchmark function $f_\text{min} = f(\mathbf{X^*})$.  We consider that a benchmark function can have more than one global minima. The relationship between these variables is represented by: $\mathbf{X^*} = \text{arg}\min_{\mathbf{X}\in \chi} f(\mathbf{X})$. 

$\mathbf{X_{\text{opt}}}$ is the output value/best solution of the optimization algorithm at the end of the budget.

$\mathbf{X_i}$ is a sampled point of the search space selected by the algorithm at function evaluation $i$.

The logged metrics are represented below 
\begin{itemize}
    \item \textbf{Final reward difference}: $\Delta f_{\text{reward}} = f(\mathbf{X_{\text{opt}}}) - f(\mathbf{X^*})$, 
    \item \textbf{Euclidean distance}: $D = \lVert \mathbf{X*} - \mathbf{X_{\text{opt}}} \rVert_2$
    \item \textbf{Solved at precision} $\epsilon \in [1, 0.1 ,1e-3, 1e-6]$: this is a boolean variable that indicates if the problem was solved or not at the end of the budget: $\Delta f_{\text{reward}} < \epsilon$
    \item \textbf{Solved at FEval}: 
    the function evaluation (FEval) in which the algorithm solved the problem with a specific precision $\Delta f_{\text{reward}} < \epsilon$
    \item \textbf{CPU time}: this computes the time spent by each algorithm in each problem for the whole budget.
\end{itemize}

The presented models often make use of a transformation of these variables, in such cases, we describe the specific transformation in each model.

\subsection{The research questions}
The presented statistical models in Section V address the following research questions.
\begin{itemize}
	\item \textbf{RQ1-a:} What is the probability of each algorithm solving a problem at precision $\epsilon \leq 0.1$? 
	\item \textbf{RQ1-b:} What is the impact of noise in the probability of success of each algorithm at precision $\epsilon \leq 0.1$?
	\item \textbf{RQ2:} What is the expected improvement of these algorithms against the Random Search in noiseless benchmark functions in terms of approaching a global minimum based on the Euclidean distance to the location of the closest global minimum?  
	\item \textbf{RQ3:} How can optimization algorithms be ranked in the conditions of 10,000 evaluations per dimension budget in noisy benchmarks?
	\item \textbf{RQ4-a:}  What is the average number of function evaluations (FEval) taken by an algorithm to converge to a solution at a precision of $\epsilon \leq 0.1$ and with a maximum budget of 100,000 function evaluations per dimension?
	\item \textbf{RQ4-b:} What is the impact of noise in the number of function evaluations (FEval)  taken by an algorithm to converge to a solution at a precision of $\epsilon\leq 0.1$ and with a maximum budget of 100,000 function evaluations per dimension?
	\item \textbf{RQ5:} Is there a difference in the CPU time taken per function evaluation between the PSO, the Random Search x1, and the Differential Evolution algorithms?
\end{itemize}

\section{Statistical Models}
\label{sec:statsmodels}
In this section, we provide an overview of five statistical models for answering different practical research questions in benchmark data. For each model, we present an introduction to the model, the model, and an analysis of the results focusing on the interpretation aspect of the intervals with plots. We conclude the discussion of each model with some final remarks, indicating possible extensions or practical issues that one may find.

It is worth reinforcing that these models are not unique and several variations can be made. Our choice for these models was based on the simplicity and ability to answer many practical questions. More complex models can be made and derived from these models. In BDA, starting with simple models and extending them is encouraged, and reporting these models can provide a greater level of transparency for research and replication studies.

Before we present the models, we provide a short overview of hierarchical/multilevel models. We emphasize the need of using hierarchical models for benchmark comparison since they can compensate for the clustering effect of the benchmarks. 

In the online appendix, we provide the empirical data used in this text, a step-by-step code used in all the data transformation, cleaning,  plots, and tables. Additionally, we provide the Stan code for all the models together with the exact data used to fit these models and the analysis of the convergence and validity of the models.

\textbf{Notation convention}:
\begin{itemize}
\item For notation clarity, we omit the indexing variable that indicates each observation of the dataset, for example, instead of $y[i] \sim \text{Normal}(a + b \cdot x[i], \sigma)$ we represent as  $y \sim \text{Normal}(a + b \cdot x, \sigma)$. Similar notations are widely used in other Bayesian data analysis texts \cite{gelman2013bayesian, mcelreath2020statistical}.
\item All dependent variables are indicated as $y$.
\item All predictors (independent) variables are indicated with $x$ with optional subscripts $i$ to indicate the algorithm.
\item All intercepts (the independent terms of the linear regression without any predictor), including the random effects of the models, are indicated by $a$ with optional subscripts $i$ to indicate the algorithm and $j$ for the benchmark.
\item All slopes of the models are indicated with $b$ with optional subscripts $i$ to indicate the algorithm.
\item The subscript index $i$ indicates that there is one parameter for each algorithm. For example, $a_i$ indicates that there is one intercept for each algorithm, $a_1$ for the first algorithm.
\item The subscript index $j$ indicates the parameter of the benchmark function. For example $a_{\text{bm},j}$ indicates that there is one intercept for each benchmark function.
\item If there is no subscript index, the parameter is common for all algorithms or benchmark functions. 
\end{itemize} 

\subsection{Compensating the effects of benchmarks}
When the measured units are drawn from the same cluster within a population (e.g. multiple samples from the same benchmark function), these can no longer be considered independent samples. This situation can add bias from unobserved variables into the model and shift the posterior distributions \cite{snijders2011multilevel}. A strategy to overcome such problems is called multilevel modeling or hierarchical modeling, and it is not restricted to the Bayesian framework. Snijders and Bosker \cite{snijders2011multilevel} present a full treatment of multilevel modeling in the frequentist setting. One approach to compensate for the clustering problem is to add a blocking variable that estimates the effect of each cluster. With a large number of clusters (e.g., 30 benchmark functions), we can model the effect of the clusters utilizing a random-effects variable. This random effect variable indicates that every benchmark will be drawn from a probability distribution (and therefore we can only estimate the parameters of this distribution), reducing model complexity and allowing us to evaluate the impact of the benchmark functions overall. Of course, it is also possible to observe the effect of each function in this framework.

In the Bayesian framework, we can condition the priors of the random effects variables over new random variables called hyperpriors. For example, let's consider the example of a simple linear regression in which each of the intercepts depends on the algorithm $a _i$ and the slope $b$ is constant for all algorithms. Variables $a$ and $b$ are not modeled as random effects and therefore are modeled only with their priors.

If we consider that each observation comes from a finite number of clusters (benchmarks), in which the cluster is represented by the index $j$, we can create the Model \ref{model:multilevelbayesian} to include a random variable intercept that represents the effect of each cluster on the observed variable. The random variable intercept for the benchmarks is represented by $a_{\text{bm},j}$. The exponential distribution is a common choice for modeling variance in random effects \cite{kruschke2013bayesian}, where lower rate parameters create proper but weakly-informative hyper prior. However, other common choices are the half-normal and the Cauchy distribution.

\model{Bayesian linear regression considering the effect of the benchmarks}{multilevelbayesian}{
    y &\sim \text{Normal}(a_i+a_{\text{bm},j}+ b\cdot x, \sigma),\\
    a_i &\sim  \text{Normal}(0,10) && \text{[Prior]},\\
    b &\sim \text{Normal}(0,10) && \text{[Prior]},\\
    \sigma &\sim \text{Exponential}(1) && \text{[Prior]},\\
    a_{\text{bm},j} &\sim  \text{Normal}(0,s) && \text{[Prior]},\\
    s &\sim \text{Exponential}(1) && \text{[Hyperprior]}.
}

In terms of interpretation, although we analyze the impact of each benchmark function (since we estimate the intercept of each benchmark function), we are more concerned with the interpretation of the standard deviation of the random effects (the $s$ parameter). This parameter indicates how much variance we can attribute to the clustering information of the benchmark functions. Additionally, suppose the selection of benchmark functions is representative of the set of problems that the algorithms are going to solve. In that case, we can interpret how much variance we can expect in the model due to a change of problem. 

The multilevel approach can be easily extended for additional levels in the hierarchy (if the function can be classified as easy or hard to solve, or other properties, such as separability or modality) and include different hierarchies in parallel. For more information regarding these extensions and other applications of multilevel models, we refer to \cite{gelman2013bayesian, mcelreath2020statistical}. Note that the separation between the main effects of the cluster effects introduces $n$ parameters in the model, in which $n$ is the number of clusters. 


\subsection{Probability of success}
In this subsection, we utilize a multilevel generalized linear model with a logit link function to model the binomial response and answer the research questions RQ1-a and RQ1-b.

\textbf{RQ1-a:} What is the probability of each algorithm solving a problem at precision $\epsilon \leq 0.1$? 

\textbf{RQ1-b:} What is the impact of noise in the probability of success of each algorithm at precision $\epsilon \leq 0.1$?

\subsubsection{The model}
One model that can be used for addressing these research questions is the generalized linear model with the binomial distribution and the inverse logit. The binomial is a common choice of the likelihood distribution in generalized linear models when one wants to estimate how many out $N$ tries are successful given a probability $p$ \cite{agresti2003categorical ,gelman2013bayesian, mcelreath2020statistical}. Through the use of generalized linear models, we can include random effects terms and predictors. This requires a link function to transform the continuous linear equation to the input of the binomial distribution. Common choices for link functions are the logit or the probit functions.  These link functions allow us to map the continuous output of the linear regression to discrete values used in the binomial distribution. The binomial model is represented by Model \ref{model:binomial}.

\model{Binomial model}{binomial}{
    y &\sim \text{Binomial}(N, p), \\
    p &= \text{logit}^{-1}(a_{\text{alg},i} + a_{\text{bm}, j} + \text{b\_noise}_i \cdot x_{\text{noise}} ) ,\\
    a_{\text{alg},i} &\sim  \text{Normal}(0,5), \\
    \text{b\_noise}_i &\sim \text{Normal}(0,5), \\
    a_{\text{bm}, j} &\sim  \text{Normal}(0,s), \\
    s &\sim \text{Exponential}(0.1). 
}

Model \ref{model:binomial} uses the following notation. Let's consider the example of one row in the dataset that indicates that the algorithm PSO was tried ten times with noise=3.0, budget=1000 for one benchmark function. On those ten tries, it solved the problem at $\epsilon=0.1$ two times. The inverse logit function, as defined below, maps the  values of $x$ from $(-\infty,+\infty)$ to the interval $(0,1)$, so its output can be used as the probability parameter of the binomial distribution
\begin{itemize}
    \item $\text{logit}^{-1}(x) = \frac{1}{1+ \exp(-x)}$.
    \item $N$: is an integer, parameter of the binomial distribution, that represents the total number of tries (in our case the aggregated value of repeated measures of a single algorithm). In the example $N=10$ .
    \item $y$: is an integer that indicates from the $N$ tries, how many of those were successful (or had a result of 1). In example, $y=2$.
    \item $a_{alg,i}$: represents the mean (intercept) effect of each algorithm. 
    \item $\text{b\_noise}_i$: is the influence of noise in each algorithm.
    \item $x_{\text{noise}}$: indicates the noise used. In the example, $x_{\text{noise}}=3.0$.
    \item $a_{bm, j}$: indicates the random effect of the benchmarks.
    \item $p$: is a variable modeled by the linear equation and it indicates the probability of success. We can use this probability to assess how the different parameters impact the probability of success.
\end{itemize}

The model above captures different coefficients for the influence of noise and budget on each algorithm. If desired (e.g. to measure the impact of factor regardless of the algorithm), these parameters could be aggregated in a single parameter.

\subsubsection{Model interpretation}
After running this model in Stan (chains=4, warmup=200, iterations=3000), we obtain the posterior distribution of the model parameters. Table  \ref{tab:probsuccesspartable} shows the mean and HPD intervals of the model parameters as well as the mean odds ratio (OR) and the OR HPD interval. The OR measure indicates the relative probability of success compared to the probability of failure. If the odds ratio is greater than 1, the parameter increases the probability of success. If the odds ratio is between 0 and 1, it decreases the probability of success.  It is worth noting, however, that if an algorithm has a high odds ratio (or parameter value) the influence of the benchmark might be small (due to the asymptotic characteristic of the inverse logit function)
\begin{table}[htb]

\caption{\label{tab:probsuccesspartable}Estimated parameters of the model. OR indicates the odds ratio of the respective parameter}
\centering
\fontsize{7}{9}\selectfont
\begin{tabular}[t]{>{\raggedright\arraybackslash}p{1.4cm}>{\raggedleft\arraybackslash}p{0.7cm}>{\raggedleft\arraybackslash}p{0.7cm}>{\raggedleft\arraybackslash}p{0.7cm}>{\raggedleft\arraybackslash}p{0.7cm}>{\raggedleft\arraybackslash}p{0.7cm}>{\raggedleft\arraybackslash}p{0.7cm}}
\toprule
Parameter & Mean & HPD low & HPD high & OR Mean & OR HPD low & OR HPD high\\
\midrule
a\_CMAES & -0.10 & -0.99 & 0.78 & 0.90 & 0.37 & 2.18\\
a\_Cuckoo & -2.55 & -3.40 & -1.63 & 0.08 & 0.03 & 0.20\\
a\_DiffEvol. & -0.21 & -1.10 & 0.66 & 0.81 & 0.33 & 1.94\\
a\_NelderM. & -4.35 & -5.24 & -3.42 & 0.01 & 0.01 & 0.03\\
a\_PSO & -0.39 & -1.26 & 0.49 & 0.67 & 0.28 & 1.63\\
a\_RandomS1 & -2.01 & -2.84 & -1.07 & 0.13 & 0.06 & 0.34\\
a\_RandomS2 & -2.22 & -3.09 & -1.33 & 0.11 & 0.05 & 0.27\\
a\_SimAnneal & -2.56 & -3.47 & -1.69 & 0.08 & 0.03 & 0.18\\
b\_CMAES & -1.27 & -1.37 & -1.18 & 0.28 & 0.26 & 0.31\\
b\_Cuckoo & -0.81 & -0.93 & -0.70 & 0.44 & 0.40 & 0.50\\
b\_DiffEvol. & -1.35 & -1.46 & -1.26 & 0.26 & 0.23 & 0.28\\
b\_NelderM. & -0.39 & -0.52 & -0.25 & 0.68 & 0.59 & 0.78\\
b\_PSO & -1.15 & -1.24 & -1.06 & 0.32 & 0.29 & 0.35\\
b\_RandomS1 & -0.97 & -1.08 & -0.86 & 0.38 & 0.34 & 0.42\\
b\_RandomS2 & -0.72 & -0.83 & -0.62 & 0.49 & 0.44 & 0.54\\
b\_SimAnneal & -0.79 & -0.91 & -0.68 & 0.45 & 0.40 & 0.51\\
s & 2.44 & 1.78 & 3.16 & 11.47 & 5.94 & 23.68\\
\bottomrule
\end{tabular}
\end{table}


Table \ref{tab:probsuccesspartable} indicates that the algorithms PSO, Differential Evolution, and CMAES have a significantly higher mean of the OR compared to the other algorithms, and are the only ones that, on average, have a higher probability of solving a problem than not solving. However, all three also have a wide OR HPD interval, meaning that their performance is greatly affected by external random factors (such as choice of seed for example). The large overlap between the OR HPD interval of those three algorithms indicates no statistical difference between them.

From the analysis of the odds ratio of the noise coefficient, we can see that all algorithms perform more poorly in the presence of noise (all the odds ratios are lower than one). Noise has a greater relative impact on the Differential Evolution, CMAES, and PSO algorithms. However, this relative effect should be analyzed in the context of the total probability of success. Since all other algorithms have a much lower probability of solving a problem, the effect of noise on them is smaller as they would probably not solve the problem even without noise. Since the logit function is not linear, it is recommended to evaluate both the marginal and absolute impact of the parameters in the model predictive accuracy. 

In terms of the effect of the benchmark functions, the effect of the benchmark (by our model) is drawn from a normal distribution with a mean of 0 and a standard deviation that has the posterior distribution of $s$. Since the effect of the benchmarks are sampled from this distribution, it can have an impact with the same magnitude or even higher than the choice of algorithms in the probability of solving a problem. This reinforces that the choice of benchmark functions greatly impacts the results of algorithms and that using higher hierarchy levels for the benchmarks may improve the estimates of the algorithms.

\subsubsection{Remarks}
The same model can be used to answer many other research questions, such as, the probability of an algorithm solving problems when duplicating the budget (such question could be used to determine an appropriated budget for some algorithms, especially in expensive functions), the probability to solve a particular problem (now the focus is on specific benchmark functions), among others. An equivalent variant of this model, one that does not use aggregated data, utilizes a Bernoulli distribution as the likelihood instead of the binomial, the inverse logit part and the priors can remain the same.

Additional discussion about categorical models (such as the Binomial and Bernoulli model) in the context of generalized linear models and possible extensions can be found in both Bayesian and frequentist statistical textbooks \cite{mcelreath2020statistical, agresti2003categorical}.

\subsection{Algorithm relative improvement over Random Search}
In this subsection, we address RQ2. For this question, we will use a model based on a linear regression. This model is analogous to the frequentist linear regression models but we have estimations of the full posterior distribution, control over the priors, and credible intervals (instead of confidence intervals) for inference. Linear models such as this one facilitate the direct comparison of the magnitude and direction of the parameters. Since we are assessing a linear model directly, interpretation mistakes over the absolute impact of transformations such as the odds-ratio do not occur.

\textbf{RQ2:} What is the expected improvement of these algorithms against the Random Search in noiseless benchmark functions in terms of approaching a global minimum based on the Euclidean distance to the location of the closest global minimum? 

\subsubsection{The model}
Model \ref{model:relativeimprovement} uses a regression model with normal distribution for the likelihood and model the standard deviation (of the error term) with a single parameter $\sigma$.

\model{Regression model}{relativeimprovement}{
    y &\sim \text{Normal}(\mu, \sigma), \\
    \mu &= a_{\text{alg},i} + a_{\text{bm}, j}, \\
    \sigma &\sim \text{Exponential}(1), \\
    a_{\text{alg},i} &\sim  \text{Normal}(0,1), \\
    a_{\text{bm}, j} &\sim  \text{Normal}(0,s), \\
    s &\sim \text{Exponential}(0.1). 
}

In Model \ref{model:relativeimprovement}, we have the following notation:
\begin{itemize}
    \item $y$: is the metric that indicates the relation between the Euclidean distance of the algorithm and the random search. There are multiple ways to measure improvement but, in this example, we will use the difference between the Random Search x1 Euclidean distance and the algorithm, divided by the Random Search x1 Euclidean distance. The results of this ratio are caped between -1 and 1. The Random Search x1 Euclidean distance is averaged in the repeated measures.
    \item $a_{alg,i}$: represents the mean (intercept) effect of each algorithm. 
    \item $a_{bm, j}$: indicates the random effect of the benchmarks.
    \item $\mu$: represents the mean value of the likelihood,  modeled by the linear equation. This is a transformation parameter and not an estimated parameter of the model. $\mu$ represents the average improvement to random search after counting for the effect of the algorithm and the benchmark but it does not include the variance added by the $\sigma$ parameter.
    \item $s$: represents the standard deviation of the effect of the benchmark functions.
\end{itemize}

\subsubsection{Model interpretation}
After running this model in Stan  (chains=4, warmup=200, iterations=2000), we get the posterior of each parameter and compute the HPD intervals for the intercept of each algorithm. The appendix provides a table with the estimates of every  posterior parameter obtained by the model. 
Table \ref{tab:relativeimprovementpartable} shows the obtained posterior parameters and their HPD intervals. The mean value on this table corresponds to the estimated parameters and not the mean value of the likelihood ($\mu$).
	\begin{table}[htb]

\caption{\label{tab:relativeimprovementpartable}Estimated parameters of the relative improvement model and their respective HPD intervals}
\centering
\fontsize{7}{9}\selectfont
\begin{tabular}[t]{lrrr}
\toprule
Parameter & Mean & HPD low & HPD high\\
\midrule
$\sigma$ & 0.64 & 0.63 & 0.65\\
a\_CMAES & 0.15 & 0.09 & 0.21\\
a\_CuckooSearch & -0.38 & -0.44 & -0.32\\
a\_DifferentialEvolution & 0.30 & 0.24 & 0.36\\
a\_NelderMead & -0.64 & -0.70 & -0.58\\
a\_PSO & 0.32 & 0.26 & 0.38\\
a\_SimulatedAnnealing & -0.57 & -0.63 & -0.51\\
s & 0.15 & 0.11 & 0.19\\
\bottomrule
\end{tabular}
\end{table}

The algorithm intercept represents the average improvement over random search in an average benchmark with a null effect. The random effect standard deviation $s$, has a similar interpretation as before. The impact of the benchmark over the relative improvement given by the benchmarks is sampled from a normal distribution with a mean equal to zero and standard deviation $s$.  The parameter $\sigma$ represents the dispersion of the results around the average $\mu$.


From Table \ref{tab:relativeimprovementpartable}, we can see that only the algorithms PSO, Differential Evolution ,and CMAES have a positive relative improvement over the Random Search x1 algorithm. All the other parameters had negative estimates, which result that they perform on average worse than random search in approaching the global minima. We can notice that the standard deviation of the measurements ($\sigma$) is high compared to the estimates of the algorithms. This indicates that there is a lot of non-explained variance in the data. For this case, since the comparison is relative between algorithms, we see that the benchmark functions have a smaller impact on the estimates, with a much lower standard deviation for the random effects $s$. 

\subsubsection{Remarks}
This model can easily be extended to include other predictors as any regression model, such as to evaluate the impact of noise and the log of budget, among others.

Often, the first regression model we try is based on the normal likelihood. However, if we see long-tail distributions or outliers, a robust regression might be more appropriate. Robust methods are discussed in our last model in this section.  The choice between the different types of regression is often subject to some experimentation to see which model works and predicts the data better. The differences between the performance of all of these models can be compared with information criteria methods such as WAIC.

It is worth noting that the MCMC is quite sensitive to numerical stability. Inferences in which data goes from wide ranges such as 0.001 to 1000 usually makes the MCMC fail to converge. This problem is particularly common in benchmark data since functions have very different ranges. One solution to this problem is to normalize the input data in respect to another baseline algorithm (which we did). Another solution commonly used approach is to compare ranking statistics. Ranking statistics have the disadvantage of throwing out part of the data. Ranking allows researchers to assess which algorithm performs better but not by how much.

\subsection{Ranking comparison}
In this subsection,  we utilize a Bayesian variant of the Bradley-Terry Model \cite{bradley1952rank, mattos2021bayesian} to answer RQ3.

\textbf{RQ3:} How can optimization algorithms be ranked in the conditions of 10,000 evaluations per dimension budget in noisy benchmarks?

\subsubsection{The model}
The Bradley-Terry model \cite{bradley1952rank} is a popular approach to investigate paired comparisons \cite{caron2012efficient}. In this model, a pair of competitors are compared and one of them is classified as the winner. This model has been used in various applications, from medicine to machine-learning applications on search engines.

The model is based on the idea of a comparison contest between players (or in our case, algorithms) without the possibility of ties. The model can be expressed through latent variables $\alpha _i$ that represent the ``strength'' or the ability of each algorithm \cite{turner2012bradley}. The odds that algorithm $i$ beats $j$ is expressed by $\alpha _i/ \alpha _j$, and the probability that $i$ beats $j$ is expressed by:

	\smallskip\noindent
	\begin{small}
	\begin{align}
	\label{eq:probbeat}
	\text{Pr}[i \text{ beats } j] &= \frac{ \exp{\alpha_i} }{  \exp{\alpha_i} +  \exp{\alpha_j}},\\
	&= \text{logit}^{-1}(\alpha_i - \alpha_j).
	\end{align}
	\end{small}

Since we do not observe the probabilities, we will model it through a Bernoulli distribution. The goal of this model is to estimate the strength of the parameters and the order of these parameters will give a rank of the algorithms. The main advantage of this model over the frequentist approach with the maximum likelihood estimator is the ability to restrict the priors and obtain a full posterior estimation of the parameters.

If we want to include other predictors as well as control for random effects in the latent strength parameter, we can use the following linear transformation for each latent strength variable  \cite{cattelan2012models}.

\begin{small}
\smallskip\noindent
\begin{equation}
\alpha_i = \sum ^p_{k=1} (\beta_k x_k) + a_{\text{bm},i, j} ,
\end{equation}
\end{small}

\noindent 
in which $p$ is the number of predictors we want to add, $\beta_k$ are the coefficients we want to estimate, $x_k$ are the independent variables for each predictor, and $a_{\text{bm},i, j}$ is the estimated effect of the random variable, such as the clustering effect of the benchmark $j$ in the latent variable $\alpha_i$. Here we note that each benchmark will provide a different effect for each strength that is estimated in the random effects variable. It is worth noting that in each paired comparison, if the same covariates are available for both contestants, they will cancel the effect on each other out.

Model \ref{model:bradleyterry} represents the Bayesian Bradley-Terry model with random effects for the benchmarks:

\model{Bayesian Bradley-Terry Model}{bradleyterry}{
   & y \sim \text{Bernoulli}(p), \\
   & p =  \text{logit}^{-1}( a_{\text{algo1}} + a_{\text{bm},\text{algo1}, j} - a_{\text{algo0}} - a_{\text{bm},\text{algo0}, j} ), \\    
   & a_i \sim \text{Normal}(0,2),\\
   & a_{\text{bm},i, j} \sim \text{Normal}(0,s),\\
   &  s \sim \text{Exponential}(0.1). 
}

In Model \ref{model:bradleyterry}, we have the following notation:
\begin{itemize}
    \item $y$: indicates which of the two algorithms (algo1 or algo0) won the contest. It can have only two values 0 for algo0 or 1 for algo1.
    \item $a_{\text{algo1}}$ and $a_{\text{algo0}}$: represents a paired comparison between two algorithms
    \item $a_i$: indicates the latent strength variable of each algorithm.
    \item $ a_{\text{bm},\text{algo0}, j}$  and  $ a_{\text{bm},\text{algo1}, j}$ are the random effects due to the benchmark $j$ in the algorithm 0 or 1.
    \item $s$: represents the standard deviation of the effect of the benchmark functions.
\end{itemize}

\subsubsection{Model interpretation}
After running this model in Stan  (chains=4, warmup=200, iterations=4000), we get the posterior of each parameter and compute the HPD intervals for the intercept of each algorithm strength. Table \ref{tab:rankingpartable} shows the summary statistics of the posterior distribution parameters of the model, the latent strength variables, and the standard deviation of the random effects of the benchmarks. 
	\begin{table}[htb]

\caption{\label{tab:rankingpartable}Estimated parameters of the ranking model and the respective HPD intervals}
\centering
\fontsize{7}{9}\selectfont
\begin{tabular}[t]{lrrr}
\toprule
Parameter & Mean & HPD low & HPD high\\
\midrule
a\_CMAES & 1.04 & -0.48 & 2.59\\
a\_CuckooSearch & -0.41 & -1.87 & 1.19\\
a\_DifferentialEvolution & 1.99 & 0.43 & 3.51\\
a\_NelderMead & -2.98 & -4.51 & -1.43\\
a\_PSO & 1.58 & 0.08 & 3.13\\
a\_RandomSearch1 & 0.28 & -1.28 & 1.78\\
a\_RandomSearch2 & 0.25 & -1.25 & 1.81\\
a\_SimulatedAnnealing & -1.69 & -3.20 & -0.11\\
s & 1.82 & 1.59 & 2.05\\
\bottomrule
\end{tabular}
\end{table}

These strength parameters can be used to either assess the probability of one algorithm beating the other or to rank the algorithms. However, despite the apparent large overlap between these latent parameters, it does not indicate that the algorithms perform similarly. To compare a specific algorithm with the other, it is possible to either compute the posterior distribution of one algorithm beating the other (as in equation \ref{eq:probbeat}) or to calculate the posterior distribution of the ranks.

To calculate the posterior distribution of the ranks, we use 1000 samples from the posterior of the strength parameters (taking into account the effect of the benchmarks) and rank them. With this procedure, we have a distribution of the ranks, which gives us also information regarding to the uncertainty of the ranking process. These results are shown in Table \ref{tab:rankingalgorithmsresults}, which shows the median rank and the rank variance for each algorithm. 
 

\begin{table}[htb]

\caption{\label{tab:rankingalgorithmsresults}Ranking the algorithms based on the reward difference taking accounting for the effect of the benchmarks}
\centering
\fontsize{7}{9}\selectfont
\begin{tabular}[t]{lrr}
\toprule
Algorithm & Median Rank & Variance of the Rank\\
\midrule
DifferentialEvolution & 1 & 0.20\\
PSO & 2 & 0.30\\
CMAES & 3 & 0.34\\
RandomSearch1 & 4 & 0.47\\
RandomSearch2 & 5 & 0.51\\
CuckooSearch & 6 & 0.21\\
SimulatedAnnealing & 7 & 0.01\\
NelderMead & 8 & 0.00\\
\bottomrule
\end{tabular}
\end{table}

\subsubsection{Remarks}
The Bradley-Terry model is the simplest model for analyzing ranks. Note that this model does not take into account the differences between the algorithms (despite how big or small they might be). We did not include any predictor in our model since the predictor variable is always the same for each algorithm comparison (since this is how we designed the experimental data collection) and they would cancel each other.

There are extensions and alternative models to the Bradley-Terry model, such as the Thurstonian model (that uses a probit instead of the logit function). The Plackett-Luce model described in \cite{calvo2019bayesian} is equivalent to the presented Bradley-Terry model when a complete rank is converted to paired comparisons with independence between the algorithms in the rank \cite{turner2020modelling}. The advantage of converting ranks into paired comparisons and using the Bradley-Terry model is the possibility to use partial ranks.

Our model does not accept ties between contestants in its formulation. Since we are comparing the true reward difference, ties only occur when the algorithms achieve the global minima, which is a rare event. One approach to solve ties is to randomly assign a winner \cite{cattelan2012models}. However, in cases where ties are common and estimating the probability of two contestant algorithms to tie is relevant, such as in combinatorial optimization, extensions to the Bradley-Terry model can be used. The most common extension to accommodate  ties is the Davidson generalization of the Bradley-Terry model \cite{davidson1970extending, mattos2021bayesian}. This extension adds an additional parameter $\nu > 0$  that estimates the overall maximum probability of a tie and the dependence of the probability of a tie in the strength parameter. If $\nu \xrightarrow{} \infty$ the probability of a draw is 1. If $\nu \xrightarrow{} 0$ the probability of tie depends only on the strength of the algorithms. In the online appendix we provide the Stan statistical model for the Davidson extension. The model adds a conditional statement in the probabilities as seen below:

\smallskip\noindent
\begin{small}
\begin{align*}
\text{Pr}[i \text{ beats } j | \text{not tie}] &= \dfrac{ \exp{\alpha_i} }{  \exp{\alpha_i} +  \exp{\alpha_j} +  \exp{(\nu +(\alpha_i + \alpha_j) / 2 ) }}, \\
\text{Pr}[i \text{ ties } j] &= \dfrac{ \exp{(\nu + (\alpha_i + \alpha_j) / 2) } }{  \exp{\alpha_i} +  \exp{\alpha_j} +  \exp{(\nu +(\alpha_i + \alpha_j) / 2 ) }} .
\end{align*}
\end{small}

\subsection{Number of function evaluations to converge to a solution}
\label{sec:timetoconverge}
In this subsection, we will consider two research questions that address the number of function evaluations (FEval) to the occurrence of an event, RQ4-a and RQ4-b. Such questions are usually discussed in the area of survival analysis, in which the primer interest is when an event occurs \cite{clark2003survival}.  

\textbf{RQ4-a:}  What is the average number of function evaluations (FEval) taken by an algorithm to converge to a solution at a precision of $\epsilon \leq 0.1$ and with a maximum budget of 100,000 FEval per dimension?

\textbf{RQ4-b:} What is the impact of noise in the number of function evaluations (FEval) taken by an algorithm to converge to a solution at a precision of $\epsilon\leq 0.1$ and with a maximum budget of 100,000 FEval per dimension?

As we noticed with the previous models, some algorithms have a very low success rate. For this example, we will use only the top 4 algorithms indicated by the previous Bradley-Terry Model: Differential Evolution, PSO, CMA-ES, and RandomSearch1. The proposed research questions address the average FEval to converge to a solution divided by the number of dimensions because we are utilizing a fixed budget per dimension as the experimental condition. This leads benchmark functions with 6 dimensions to be evaluated up to 3 times higher than benchmark functions with only 2 dimensions. Without correcting the number of FEval by the dimensions, the number of dimensions in a benchmark function would create a higher difference than the choice of algorithms. This transformation ensures that the results of different benchmarks are comparable regardless of the number of dimensions.

\subsubsection{The model}
One important aspect of survival models, that makes them different from other analyses, is the presence of censored data. Censored refers to the characteristic that an event might not occur in the window of observation. If we do not consider censoring, we will be eventually creating a downwards bias in our inference \cite{mcelreath2020statistical}. Although there are many ways that data can be censored, in the analysis of benchmark functions, we are concerned primarily with uninformative and \textit{right censoring}. Uninformative, because the censored data does not contain information regarding the survival (e.g. the data is censored because the algorithm has a bug and never converges to a solution) and right censor because we do not observe the event due to the end of the window of observation \cite{clark2003survival}. In our example, right censoring is when an algorithm is unable to solve the problem given the budget. 

Survival analysis is usually modeled in terms of two related probability functions, the survival and the hazard functions. The survival function $S(t)$ models that an algorithm will not converge until function evaluation $t$. The hazard function $h(t)$ models the probability that an algorithm will converge at function evaluation $t$ during that particular evaluation, i.e., it is the instantaneous event rate of an algorithm converging to a solution if it hasn't yet. More commonly, we use the cumulative hazard function $H(t)$, obtained by integrating $h(t)$ over the number of FEval. In summary, the hazard function models the occurrence of the event (converge to a solution) and the survival function models the non-occurrence of the event. The relationship between $S(t)$ and $H(t)$ is given by: $H(t) = -\log S(t).$

We will use the most common (and simple) survival model, the Cox's Proportional Hazard model and the Bayesian formulation proposed by Kelter \cite{kelter2020bayesian} for the random effects. This model assumes a time-invariant exponential hazard function and is easily extended with additional predictors. In this model, the hazard function is constant in time and we refer to it as $\lambda(\mathbf{X})$, in which $\mathbf{X}$ is the matrix of covariates, $a$ is a constant baseline hazard (if all covariates are zero) and $\mathbf{b}$ is the corresponding matrix of the coefficient of the covariates. The expected value for the occurrence of the event is the inverse of the hazard function and defined as:

\begin{small}
\begin{subequations}
\label{eq:hazardrate}
\begin{align}
\SwapAboveDisplaySkip
    h(t) &= \lambda(\mathbf{X}) = \exp(a+\mathbf{b X}) ,\\
    \mu (\mathbf{X}) &= \frac{1}{\lambda(\mathbf{X})}.
\end{align}
\end{subequations}
\end{small}

In the Bayesian framework (instead of using the partial likelihood method), we divide the model into censored and non-censored parts. While the non-censored data uses the hazard function above, the censored data uses the complementary cumulative probability distribution function. Our model is represented by Model \ref{model:coxhazard}:

\model{Bayesian Cox's Proportional Hazard}{coxhazard}{
&\text{If event}=1: \\
&y \sim \text{Exponential}(\lambda_{i,j} ) \\ 
&\text{If event}=0: \\
&y \sim \text{Exponential-CCDF}(\lambda_{i,j} ) \\
&\lambda_{i,j} =  \exp{(a_{\text{alg},i} + a_{\text{bm}, j}}  + b_{\text{noise},i}\cdot x_\text{noise}), \\
&\mu_{i,j} = \frac{1}{\lambda_{i,j} },\\
&\text{Priors}:\\
&b_{\text{noise},i} \sim \text{Normal}(0,2),\\
&a_{\text{alg},i} \sim \text{Normal}(0,10), \\
&a_{\text{bm}, j} \sim  \text{Normal}(0,s), \\
&s \sim \text{Exponential}(0.1).
}

In Model \ref{model:coxhazard}, we have the following notation:
\begin{itemize}
	\item event: is a binary variable that indicates if the algorithm has found a solution or not.
    \item $y$: if event=1, $y$ represents the function evaluation that the algorithm finds a solution divided by the number of dimensions. If event=0, $y$ is considered a missing value.
    \item $a_{\text{alg},i}$ represents the baseline effect of each algorithm. 
    \item $b_{\text{noise},i}$: is the influence of noise in the FEval to find a solution of the algorithm.
    \item $x_\text{noise}$: indicates the noise of the benchmark function.
    \item $a_{\text{bm}, j}$: indicates the baseline of the random effects of the benchmarks.
\end{itemize}

\subsubsection{Model interpretation}
After running this model in Stan  (chains=4, warmup=200, iterations=3000), we get the posterior of each parameter and compute the HPD intervals for each parameter. Table \ref{tab:timetoconvergepartable} shows the obtained posterior parameters and the HPD intervals. Combined with equations 5a and 5b, we can infer that the algorithms that have a lower baseline effect have a higher probability of surviving, i.e., not finding a solution in the specified budget. The average FEval to converge in the noiseless case can be seen in table \ref{tab:averagetimetoconverge_hr_table}. The presence of a random noise also increases the survival probability, as expected, and the lower the value of the noise variable, the more it impacts the ability of the algorithm to find a solution.

\begin{table}[htb]

\caption{\label{tab:timetoconvergepartable}Estimated parameters of the time to converge model and the respective HPD intervals}
\centering
\fontsize{7}{9}\selectfont
\begin{tabular}[t]{lrrr}
\toprule
Parameter & Mean & HPD low & HPD high\\
\midrule
a\_CMAES & -5.09 & -6.03 & -4.16\\
a\_DifferentialEvolution & -6.37 & -7.29 & -5.42\\
a\_PSO & -6.30 & -7.22 & -5.36\\
a\_RandomSearch2 & -8.95 & -9.91 & -8.02\\
b\_CMAES & -0.79 & -0.93 & -0.66\\
b\_DifferentialEvolution & -0.96 & -1.09 & -0.83\\
b\_PSO & -0.68 & -0.80 & -0.56\\
b\_RandomSearch2 & -0.41 & -0.55 & -0.27\\
s & 2.44 & 1.85 & 3.12\\
\bottomrule
\end{tabular}
\end{table}

The average number of FEval taken by an algorithm to converge (RQ4-a) is based on the expected value of the exponential distribution which is given by the $\mu_{i,j}$ variable. By sampling 1000 values of the posterior distribution of the Cox's hazard model, we can compute the average function evaluation to converge for any experiment condition or benchmark. Table \ref{tab:averagetimetoconverge_hr_table} shows the average FEval to converge and the HPD intervals for the noiseless condition and the average of the benchmark functions (the condition where $a_{\text{bm}, j}=0$). We can see that in these conditions CMAES has a lower average number of function evaluations to converge compared to the others, while both PSO and Differential Evolution have approximately the same interval range. Note that these intervals are relatively wide due to the diversity of the benchmark functions and the uncertainty added by right censoring (when an algorithm is not able to find a solution in the determined budget). To investigate the average for each benchmark or for the scenario with noise, we can substitute the equivalent estimated parameters and predictors in the equations of the model \ref{model:coxhazard} to obtain the $\mu_{i,j}$. A computational example of this procedure is shown in the appendix.

To facilitate the interpretation of the Cox's regression model, we can also analyze the hazard ratio (HR) quantities and the baseline hazard $h_0 = \exp(a_{\text{alg},i})$. The HR represents the contribution of a parameter in the probability of occurring an event (solving the benchmark problem). The HR is defined as $\text{HR}(b)=\exp(b)$, if HR is greater than one the parameter increases the chance of the occurrence of the event, if it is less than 1, it reduces the chance of the occurrence of the event. Table \ref{tab:averagetimetoconverge_hr_table} shows the HR of the mean value of the parameters. We can see that all algorithms have a low baseline for the hazard and that noise reduces this hazard even further, therefore, increasing the number of function evaluations to converge to a solution. We can notice as well that noise impacts the hazard of the Differential Evolution algorithm more and the RandomSearch1 much less.

\begin{table}[htb]

\caption{\label{tab:averagetimetoconverge_hr_table}Average FEval to converge and the Hazard Ratio for the FEval to converge model}
\centering
\fontsize{7}{9}\selectfont
\begin{tabular}[t]{lrrrrr}
\toprule
\multicolumn{1}{c}{ } & \multicolumn{3}{c}{Avg FEval} & \multicolumn{2}{c}{Hazard Ratio} \\
\cmidrule(l{3pt}r{3pt}){2-4} \cmidrule(l{3pt}r{3pt}){5-6}
Parameter & Mean & HPD low & HPD high & Baseline & Noise\\
\midrule
CMAES & 179 & 42 & 349 & 0.006 & 0.456\\
DifferentialEvolution & 663 & 177 & 1292 & 0.002 & 0.386\\
PSO & 601 & 142 & 1159 & 0.002 & 0.509\\
RandomSearch1 & 6663 & 1746 & 13252 & 0.000 & 0.561\\
\bottomrule
\end{tabular}
\end{table}

\subsubsection{Remarks}
In the model of survival data with the Cox regression, it is important to add the random effects of the benchmark functions. Since the algorithms often cannot solve a problem regardless of the budget, if we do not include the effects of the benchmarks, we underestimate the hazard ratio of the algorithms.


The proposed model assumes that the number of function evaluations in which the algorithms converge is independent of the number of dimensions of benchmark function and therefore this effect is included in the random term. However, if the set of benchmarks includes multiples times the same function with different dimensions and the researcher wants to investigate the effect of the number of dimensions in the number of function evaluations to converge, the number of dimensions can be added as a linear predictor to the Cox hazard model similar to how it was conducted with the effect of noise.

In our case, we assume survival functions based on the exponential distribution, however, often different likelihoods such as the Weibull, Gamma, and Log-Normal distribution can provide a better fit if the predictive accuracy of the exponential model is low.

\subsection{Multiple group comparison of CPU time}
This last model addresses the specific problem of robust multiple group comparisons. We consider RQ5:

\textbf{RQ5:} Is there a difference in the CPU time taken per function evaluation between the PSO, the Random Search x1, and the Differential Evolution algorithms?

\begin{figure}[htb]
\centering
\includegraphics[width=0.35 \textwidth]{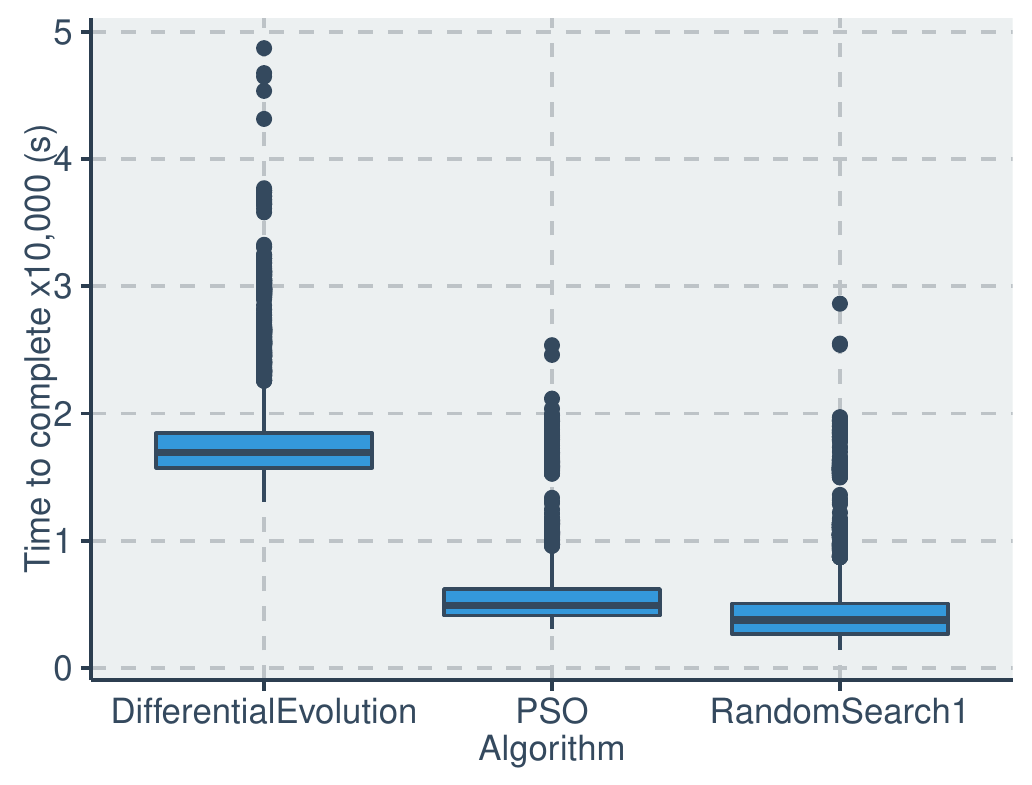}
\caption{Box-plot of the CPU time to complete of the three algorithms}
\label{fig:groupcomparisonexplore}
\end{figure}

The motivation for this question comes from an exploratory visual analysis from Figure \ref{fig:groupcomparisonexplore}. From this figure, it is clear that the Differential Evolution is slower than the others but although the box-plot suggests that RandomSearch1 is faster than the PSO, we have multiple outliers and heavy-tail distributions in the data. Besides the discussed problems in frequentist analysis (including the impossibility to accept the null hypothesis), we cannot perform a frequentist regression or ANOVA, because the residuals are not normally distributed and the homoscedasticity assumption is not met \cite{kruschke2013bayesian}. The CPU time per function evaluation is not exponentially distributed and transformations such as the log of the CPU time still present heavy-tailed distributions (additional information such as plots to support this statement are available in the appendix).

\subsubsection{The model}
This model is an extension of the BEST (Bayesian Estimation Supersedes the t-Test) approach presented by Kruschke \cite{kruschke2013bayesian} with added pooled random effects for each group. Each algorithm is modeled individually utilizing a Student-T distribution, and assuming that there is not homogeneity of the variances between the algorithms. In the Student-T distribution, we estimate a single degree of freedom $\nu$ parameter for all distributions. The model is represented by Model \ref{model:bestmodel}.

\model{Robust multiple comparison model}{bestmodel}{
    y &\sim \text{Student-T}(\nu, \mu_i, \sigma_i), \\
    \mu_i &= a_{\text{alg},i} + a_{\text{bm}, j}, \\
    \sigma_i &\sim \text{Exponential}(1), \\
    \nu &\sim \text{Exponential}(1/30), \\
    a_{\text{alg},i} &\sim  \text{Normal}(0,1), \\
    a_{\text{bm}, j} &\sim  \text{Normal}(0,s), \\
    s &\sim \text{Exponential}(1).
}

In Model \ref{model:bestmodel}, we have the following notation:
\begin{itemize}
    \item $y$: is the metric that indicates the CPU time to complete one function evaluation. We compute the total CPU time of the optimization including the CPU time spent on evaluating the benchmark function and divide it by the number of function evaluations multiplied by 10,000.
    \item $a_{alg,i}$: represents the mean (intercept) effect of each algorithm. 
    \item $\sigma_i$: indicates the standard deviation of the Student-T distribution of each algorithm.
    \item $a_{bm, j}$: indicates the random effect of the benchmarks.
    \item $\mu_i$: represents the mean value of each algorithm in the linear equation.
    \item $s$: represents the standard deviation of the effect of the benchmark functions.
    \item $\nu$: represents the degrees of freedom of the Student-t distribution modeling the tails of the distribution for the robustness. We initialize it with a long tail prior.
\end{itemize}


\subsubsection{Model interpretation}
After running this model in Stan  (chains=4, warmup=200, iterations=3000), we get the posterior of each parameter and compute the HPD intervals for the intercept of each algorithm and its standard deviation parameter.
Table \ref{tab:multiplegroupsdifferenceartable} shows the obtained posterior parameters of the model and the HPD intervals. 
	\begin{table}[htb]

\caption{\label{tab:multiplegroupsdifferenceartable}Estimated parameters of the multiple group comparison model and the respective HPD intervals}
\centering
\fontsize{7}{9}\selectfont
\begin{tabular}[t]{lrrr}
\toprule
Parameter & Mean & HPD low & HPD high\\
\midrule
a\_DifferentialEvolution & 1.78 & 1.66 & 1.90\\
a\_PSO & 0.57 & 0.45 & 0.69\\
a\_RandomSearch1 & 0.44 & 0.32 & 0.56\\
$\sigma$\_DifferentialEvolution & 0.09 & 0.08 & 0.09\\
$\sigma$\_PSO & 0.07 & 0.07 & 0.07\\
$\sigma$\_RandomSearch1 & 0.04 & 0.04 & 0.04\\
s & 0.31 & 0.23 & 0.39\\
$\nu$ & 2.75 & 2.59 & 2.90\\
\bottomrule
\end{tabular}
\end{table}



The degrees of freedom parameter $\nu$ is low, indicating that when we estimate the model, the data indeed have long tails, which reinforces the need for a robust regression. If $\nu > 30$, the  Student-T distribution approaches a normal distribution and indicates that the model could also be modeled with similar results by a normal distribution and that the presence of outliers did not impact much the estimation of the parameters. Additionally, $\sigma_i$ parameters indicate that, due to the non-overlapping intervals, there is no homogeneity of variances, which prevents the use of the traditional ANOVA for the multiple group comparison (as it is a pre-requisite of many familywise comparison methods). The effect of the benchmarks can be estimated from a normal distribution with a mean equal to zero and with a standard deviation equal to the posterior distribution of $s$. The intercept parameters of the benchmarks drawn from this distribution indicate which functions introduce or reduce the CPU time to the completion of the algorithms compared to the average CPU time.

Since we want to estimate the difference between the pairs of algorithms, the difference can be calculated by sampling (in this case 10000 times) the posterior distribution of the intercepts of the algorithms and calculating the difference. This results in a new posterior distribution of the differences. Table \ref{tab:multiplegroupsdifference} shows the HPD intervals of the differences between groups. The non-zero overlapping HPD intervals indicate a real difference between the CPU time to the completion of each of these algorithms.

\begin{table}[htb]

\caption{\label{tab:multiplegroupsdifference}HPD interval for the difference between the groups}
\centering
\fontsize{7}{9}\selectfont
\begin{tabular}[t]{lrrr}
\toprule
Difference & Mean & HPD low & HPD higher\\
\midrule
PSO - RandomSearch & 0.13 & 0.12 & 0.13\\
DiffEvolution - PSO & 1.21 & 1.21 & 1.22\\
DiffEvolution - RandomSearch & 1.34 & 1.33 & 1.34\\
\bottomrule
\end{tabular}
\end{table}

\subsubsection{Remarks}
Other distributions can be used instead of the Student-T distribution for other types of robust regression, such as a double exponential distribution \cite{gelman2013bayesian}. This robust multiple comparison model can also be easily extended to incorporate other predictors in the linear regression of each algorithm, which is not possible in non-parametric frequentist models. 


\subsection{Extending the models}
The presented models in this section can answer a variety of research questions. However, different problems, research questions, and experimental conditions might require modifications and extensions that go beyond the proposed models. In the remarks subsection of each model, we discuss possible extensions specific to those models. The proposed models are aimed at being the first step into the BDA. We reinforce that the models we presented, despite the simplicity of being based on the linear regression, still address the clustering information from benchmarks and can take other predictors into account.  Models based on linear regression are useful to answer questions about the direction and the magnitude of the effect of independent variables \cite{furia2019bayesian}. 

Different experimental conditions, as well as new covariates, can be controlled and their effects investigated by adding the predictor terms in the linear equation, similar to how the noise covariate was compensated in the binomial and in the Cox's Proportional Hazard models. Extending the model with transformation in the covariates, for example adding the log of the maximum budget, and investigating the effects of interactions as well as adding higher-order predictors (if the relationship with the predictor is not linear) are also possible \cite{gelman2013bayesian, mcelreath2020statistical}.

The random effects models we used only take into account the repeated measures of the benchmarks. However, higher levels can be introduced to investigate other effects \cite{snijders2011multilevel, gelman2013bayesian} (such as the difficulty level, if it is separable or not, etc). Adding additional clustering information follows a similar procedure as presented at the beginning of this section. Continuous random effects (instead of categorical variables) are also possible through the usage of Gaussian or Dirichlet processes as priors and are often discussed in Bayesian hierarchical modeling textbooks \cite{gelman2013bayesian, mcelreath2020statistical}.

As mentioned in the model comparison discussion (\ref{sec:modelcomparison}), one important aspect of BDA is the comparison of different valid model candidates. A recommended approach is to start building complex models from simple ones, such as the ones presented. If the new complex model indeed increases predictive accuracy and reduces information entropy, then the complex model is more adequate. Sensitivity analysis (\ref{sec:sensitivity}) is also valuable to analyze how much the conclusions of a new complex model diverge from simple ones and why it happens. If the conclusions do not diverge, it increases the confidence that the results are not specific to the proposed model.
\section{Conclusion}
\label{sec:conclusion}

Bayesian Data Analysis (BDA) can address many of the shortcomings of the traditional frequentist analysis and provides a greater level of flexibility in the modeling and the transparency of the model assumptions. However, the use of BDA is not widely spread in the analysis of empirical data in the evolutionary computing community.

With this paper, we argue for the adoption of BDA and present related concepts to ensure the validity of the models. We then present and discuss a set of five Bayesian statistical models that are capable of addressing a range of different research questions, that can be easily extended for new covariates and that take into account the clustering information that the benchmark functions introduce in the results by making use of multilevel models.

\section*{Acknowledgment}
This work was partially supported by the Wallenberg Artificial Intelligence, Autonomous Systems and Software Program (WASP) funded by the Knut and Alice Wallenberg Foundation, by the Software Center and with support from Google Research Cloud Credits for Ph.D. candidates. The authors would like to thank the comments made by Erika M. S. Ramos.

\bibliographystyle{./bib/IEEEtran}
\bibliography{./bib/bibliography}
\vspace{-1 cm}
\begin{IEEEbiographynophoto}{David Issa Mattos}
is a Ph.D. candidate in Software Engineering in the
Department of Computer Science and Engineering at Chalmers University of Technology. 
\end{IEEEbiographynophoto}
\vspace{-1 cm}
\begin{IEEEbiographynophoto}{Jan Bosch}
is a professor at Chalmers University Technology in Gothenburg, Sweden. 
\end{IEEEbiographynophoto}
\vspace{-1 cm}
\begin{IEEEbiographynophoto}{Helena Holmström Olsson}
is a professor in Computer Science at Malmö University in Sweden. 
\end{IEEEbiographynophoto}
\end{document}